\documentclass[prd,aps,reprint,preprintnumbers,superscriptaddress,floatfix,flushbottom,nobibnotes,nofootinbib,a4paper,showpacs,amsmath]{revtex4-1}
\usepackage{amssymb,dsfont}
\usepackage{graphicx}
\begin{document}
\title
{The strange and light quark contributions to the nucleon mass
from Lattice QCD} 

\author{Gunnar S.~Bali}\email{gunnar.bali@ur.de}
\affiliation{Institut f\"ur Theoretische Physik, Universit\"at Regensburg, 93040 Regensburg, Germany}
\author{Sara~Collins}\email{sara.collins@physik.uni-regensburg.de}
\affiliation{Institut f\"ur Theoretische Physik, Universit\"at Regensburg, 93040 Regensburg, Germany}
\author{Meinulf~G\"{o}ckeler}        
\affiliation{Institut f\"ur Theoretische Physik, Universit\"at Regensburg, 93040 Regensburg, Germany}
\author{Roger~Horsley}  
\affiliation{School of Physics, University of Edinburgh, Edinburgh EH9 3JZ, UK}
\author{Yoshifumi~Nakamura}  
\affiliation{RIKEN Advanced Institute for Computational Science, Kobe, Hyogo 650-0047, Japan}
\author{Andrea~Nobile}  
\affiliation{Institut f\"ur Theoretische Physik, Universit\"at Regensburg, 93040 Regensburg, Germany}
\author{Dirk~Pleiter}  
\affiliation{JSC, Research Center J\"{u}lich, 52425 J\"{u}lich, Germany}
\affiliation{Institut f\"ur Theoretische Physik, Universit\"at Regensburg, 93040 Regensburg, Germany} 
\author{P.E.L.~Rakow}  
\affiliation{Theoretical Physics Division, Department of Mathematical Sciences, University of Liverpool, Liverpool L69 3BX, UK}
\author{Andreas Sch\"{a}fer}  
\affiliation{Institut f\"ur Theoretische Physik, Universit\"at Regensburg, 93040 Regensburg, Germany}  
\author{Gerrit Schierholz}  
\affiliation{Deutsches Elektronen-Synchrotron DESY, 22603 Hamburg, Germany}
\author{Andr\'e Sternbeck}
\affiliation{Institut f\"ur Theoretische Physik, Universit\"at Regensburg, 93040 Regensburg, Germany}
\author{James~M.~Zanotti} 
\affiliation{Special Research Centre for the Subatomic Structure of Matter, School of Chemistry \& Physics, University of Adelaide, South Australia 5005, Australia}
\affiliation{School of Physics, University of Edinburgh, Edinburgh EH9 3JZ, UK}

\collaboration{QCDSF Collaboration}
\date{\today}
\begin{abstract}
We determine the strangeness and light quark fractions of
the nucleon mass by computing the quark line connected and
disconnected contributions to the
matrix elements $m_q\langle N|\bar{q}q|N\rangle$
in lattice QCD, using the non-perturbatively improved
Sheikholeslami-Wohlert Wilson Fermionic action.
We simulate $n_{\mathrm{F}}=2$ mass degenerate sea quarks 
with a pion mass of about 285~MeV and a lattice
spacing $a\approx 0.073$~fm.
The renormalization of the matrix elements
involves
mixing between contributions from different quark flavours.
The pion-nucleon $\sigma$-term is extrapolated to
physical quark masses exploiting the sea quark mass dependence
of the nucleon mass.
We obtain the renormalized values $\sigma_{\pi\mathrm{N}}=(38\pm 12)$ MeV
at the physical point
and $f_{T_s}=\sigma_s/m_{\mathrm{N}}=
0.012(14)^{+10}_{-3}$ for the strangeness contribution at our larger
than physical sea quark mass.
\end{abstract}
\preprint{Adelaide ADP-11-33/T755, Edinburgh 2011/33, Liverpool LTH 930}
\maketitle
\section{Introduction}
Most of the nucleon's mass is 
generated by the spontaneous breaking of chiral symmetry
and only a small part can be attributed
directly to the masses of its valence and sea quarks.
The quantities
\begin{equation}
f_{T_q}=m_q\langle N|\bar{q}{q}|N\rangle/m_{\mathrm{N}}
\end{equation}
parameterize the fractions
of the nucleon mass $m_{\mathrm{N}}$ that are carried by
quarks of flavour $q$.
Almost all visible matter of the universe is composed of
nucleons and yet little is known experimentally about
these quark contributions to their mass.

The scalar matrix elements
\begin{equation}
\sigma_q=m_q\langle N|\bar{q}{q}|N\rangle=m_{\mathrm{N}}f_{T_q}
\end{equation}
also determine the coupling strength of the
Standard Model (SM) Higgs boson (or of any similar scalar particle)
at zero recoil to the nucleon.
This then might couple to heavy particles that
could be discovered in LHC experiments, some of
which are dark matter candidates~\cite{Ellis:2009ai}.
The combination $m_{\mathrm{N}}\sum_qf_{T_q}$, $q\in\{u,d,s\}$,
will appear quadratically in this cross section
that is proportional to $|f_{\mathrm{N}}|^2$ where
\begin{equation}
\label{eq:fn}
\frac{f_{\mathrm{N}}}{m_{\mathrm{N}}}=\sum_{q\in\{u,d,s\}}\!\!\!\!\!\!f_{T_q}\frac{\alpha_q}{m_q}
+\frac{2}{27}f_{T_G}\!\!\!\!\!\!\sum_{q\in \{c,b,t\}}\!\!\frac{\alpha_q}{m_q}\,,\end{equation}
with the couplings
$\alpha_q\propto m_q/m_W$. Due to the trace anomaly
of the energy momentum tensor one obtains~\cite{Shifman:1978zn}
\begin{equation}
\label{eq:ftg}
f_{T_G}= 1-\!\!\!\!\sum_{q\in\{u,d,s\}}\!\!\!\!f_{T_q}\,.
\end{equation}
Note that the coupling $f_{\mathrm{N}}$ of Eq.~(\ref{eq:fn}) only
mildly depends on the masses
of heavy quark
flavours~\cite{Shifman:1978zn,Kryjevski:2003mh}.

The $\sigma_q$-terms
are also needed
for precision measurements of SM parameters in $pp$ collisions
at the LHC. For instance the
resolution of a (hypothetical) mass difference between
the $W^+$ and $W^-$ bosons is limited by our knowledge
of the asymmetries between the up and down as well as between the strange and
charm sea quark contents of the proton~\cite{Aad:2009wy}.
An accurate calculation of
these quantities will help to increase
the precision of SM phenomenology
and to shed light on non-SM processes.

The light quark contribution, the pion-nucleon $\sigma$-term, is defined as
\begin{align}
\sigma_{\pi\mathrm{N}}&=\sigma_u+\sigma_d=
m_u\frac{\partial m_{\mathrm{N}}}{\partial m_u}
+m_d\frac{\partial m_{\mathrm{N}}}{\partial m_d}\\\nonumber&\approx
\left.m^2_{\mathrm{PS}}\frac{dm_{\mathrm{N}}}{dm_{\mathrm{PS}}^2}\right|_{m_{\mathrm{PS}}=m_{\pi}}
\!\!\!\!\!,
\end{align}
where $m_{\mathrm{PS}}$ denotes the pseudoscalar mass.
Some time ago, employing dispersive analyses of pion-nucleon
scattering data,
the values~\cite{Gasser:1990ce}
$\sigma_{\pi N}=45(8)$~MeV
and~\cite{Pavan:2001wz} $\sigma_{\pi\mathrm{N}}=64(7)$~MeV were obtained
while a calculation in the framework of
$\mathcal{O}(p^4)$ heavy baryon chiral perturbation
theory resulted in~\cite{Borasoy:1995zx} $\sigma_{\pi\mathrm{N}}=48(10)$~MeV.
A recent covariant baryon chiral perturbation theory
(B$\chi$PT) analysis of the available pion-nucleon
scattering data~\cite{Alarcon:2011zs} resulted in the value 
$\sigma_{\pi\mathrm{N}}=59(7)$~MeV.
Even less is known about the strangeness contribution
$\sigma_s$. Since no elastic Higgs-nucleon scattering
experiments exist all phenomenological estimates largely
depend on modelling. Therefore input from lattice simulations
is urgently required.

We have witnessed an upsurge of interest in
calculating flavour singlet matrix elements recently, either
directly~\cite{Babich:2010at,Doi:2010nw,Alexandrou:2011ar,Engelhardt:2010zr,Takeda:2010cw},
by calculating the corresponding quark line connected and
disconnected terms, or indirectly, via the Feynman-Hellmann
theorem~\cite{Durr:2011mp,roger,Takeda:2010cw,Horsley:2010th,Toussaint:2009pz,Young:2009zb,Young:2009ps}.
High statistics simulations including light sea quarks
mean that reasonable signals can be obtained for
disconnected terms. Similarly, the small statistical uncertainty on
baryon masses as functions of the quark masses enable reasonable
fits to be made. Ideally, the results of both approaches should agree.

Preliminary results of this study were presented at
past Lattice conferences~\cite{Bali:2009dz,Collins:2010gr}.
This article is organised as follows. In Sec.~\ref{sec:method}
we detail the gauge configurations, simulation parameters and methods used.
In Sec.~\ref{sec:renorm} we then explain how the lattice results
are renormalized and finally we present our results in Sec.~\ref{sec:results},
before we summarize.

\section{Simulation Details and Methods}
\label{sec:method}
We simulate $n_{\mathrm{F}}=2$ non-perturbatively
improved Sheikholeslami-Wohlert fermions, using the
Wilson gauge action, at $\beta=5.29$
and $\kappa=\kappa_{ud}=0.13632$. 
Setting the
scale from the chirally extrapolated nucleon mass,
we obtain~\cite{sternbeck} the value $r_0=0.508(13)$~fm
for the Sommer scale, in the physical
limit. This results
in the lattice spacing
\begin{equation}
a^{-1}=(6.983\pm 0.049)\,r_0^{-1}=(2.71\pm 0.02\pm 0.07)\,\mathrm{GeV}\,,
\end{equation}
where the
errors are statistical and from the scale setting, respectively.
The $r_0/a$ ratio is obtained by chirally extrapolating
the QCDSF $\beta=5.29$ simulation points~\cite{najjar}.
An extrapolation of the axial Takahashi-Ward identity (AWI) mass
yields the critical hopping parameter
value
\begin{equation}
\kappa_{\mathrm{c,sea}}=0.1364396(84)\,.
\end{equation}

\begin{figure}[t]
\centerline{
\includegraphics[width=.36\textwidth,clip=]{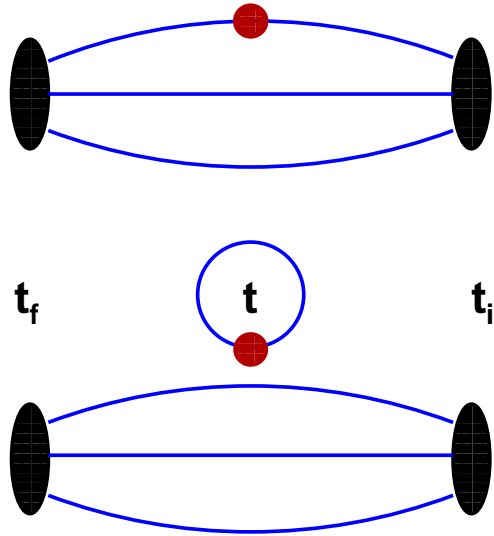}}
\caption{Quark line connected (top) and disconnected (bottom)
three-point functions. We have omitted the relative
minus sign between the diagrams. Note that for scalar matrix elements,
the vacuum expectation value of the
current insertion needs to be subtracted
($\bar{q}q\mapsto\bar{q}q-\langle\bar{q}q\rangle$), see
Eq.~(\protect\ref{eq:rati}).\label{fig:ratio}}
\end{figure}

In addition to $\kappa_{ud}=\kappa_{\mathrm{sea}}=0.13632$,
we realize the valence $\kappa$-values,
$\kappa_m=0.13609$ and $\kappa_s=0.13550$. The corresponding three
pseudoscalar masses read
\begin{align}
\label{eq:mass1}
m_{\mathrm{PS},ud}&=(0.1050\pm 0.0003)\,a^{-1}\nonumber\\
&=(285\pm 3\pm 7)\,\mathrm{MeV}\,,\\
m_{\mathrm{PS},m}&=(449\pm 3\pm 11)\,\mathrm{MeV}\,,\\
m_{\mathrm{PS},s}&=(720\pm 5\pm 18)\,\mathrm{MeV}\,.\label{eq:mass3}
\end{align}
The strange quark mass was fixed so that the above
value for $m_{\mathrm{PS},s}$ is close to the mass of
a hypothetical strange-antistrange pseudoscalar meson:
$(m_{K^{\pm}}^2+m_{K^0}^2-m_{\pi^{\pm}}^2)^{1/2}\approx
686.9$~MeV.
We investigate volumes of
$32^364$ and $40^364$ lattice points,
i.e., $Lm_{\mathrm{PS}}= 3.36$ and 4.20,
respectively, where the largest spatial lattice extent is
$L\approx 2.91$~fm. We analyse 2024 thermalized trajectories on
each of the volumes. To effectively eliminate autocorrelations for
the observables that we are interested in, bin sizes of eight are
found to be sufficient.

The matrix element
$\langle N|\bar{q}q|N\rangle$
is extracted from the ratio of
three-point functions, see
Fig.~\ref{fig:ratio},
to two-point functions at zero momentum.
Defining
\begin{equation}
\mathrm{Tr}_t A=\sum_{\mathbf{x}}
\mathrm{Tr}\,A(\mathbf{x},t;\mathbf{x},t)\,,
\end{equation}
we can write
the disconnected part as
\begin{equation}
\label{eq:rati}
R^{\mathrm{dis}}(t_{\mathrm{f}},t) = 
\left\langle\mathrm{Tr}_t(M^{-1}\mathds{1})
\right\rangle-\frac{
\left\langle
C_{\mathrm{2pt}}(t_{\mathrm{f}})
\mathrm{Tr}_t(M^{-1}\mathds{1})\right\rangle}
{\left\langle C_{\mathrm{2pt}}(t_{\mathrm{f}})\right\rangle}\,,
\end{equation}
where $M$ is the lattice Dirac operator for the quark
flavour $q$ of the current
and $C_{\mathrm{2pt}}(t_{\mathrm{f}})$ denotes the two-point function 
of the zero momentum projected proton connecting the source time
$t_{\mathrm{i}}=0$ with $t_{\mathrm{f}}$. 
Note that, unlike its expectation value, $C_{\mathrm{2pt}}$ computed on
one configuration will in general also have an imaginary part. 
This means that we can reduce the variance of the above expression by
explicitly setting this to zero, using the relation
$\mathrm{Im}\,\mathrm{Tr}_t(M^{-1}\mathds{1})=0$ that follows
from the $\gamma_5$-Hermiticity, $M^{\dagger}=\gamma_5M\gamma_5$.

\begin{figure}
\includegraphics[height=.48\textwidth,angle=270,clip=]{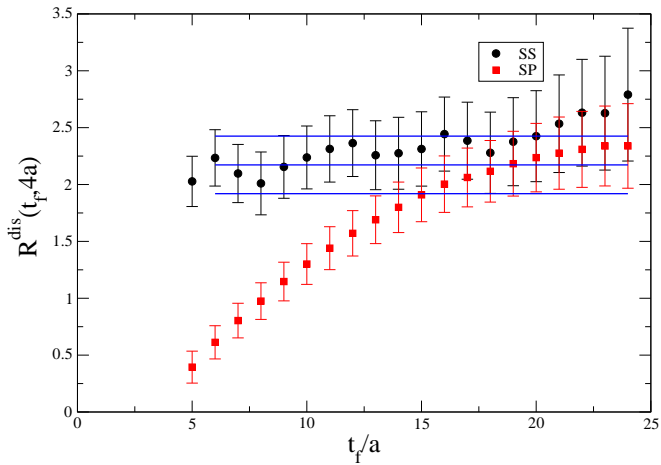}
\caption{Dependence of $R^{\mathrm{dis}}$ on $t_{\mathrm{f}}$
for smeared-smeared (SS) and smeared-point (SP) two-point functions,
together with the fit result.
\label{fig:ratio2}}
\end{figure}

In the limit of large times, $t_{\mathrm{f}}\gg t\gg 0$,
\begin{equation}
\label{eq:tlimit}
R^{\mathrm{dis}}(t_{\mathrm{f}},t)+R^{\mathrm{con}}(t_{\mathrm{f}},t) \longrightarrow
\langle N|\bar{q}q|N\rangle\,,
\end{equation}
where the term
$R^{\mathrm{con}}(t_{\mathrm{f}},t)=C_{\mathrm{3pt}}(t_{\mathrm{f}},t)/C_{\mathrm{2pt}}(t_{\mathrm{f}})$
only contributes for $q\in\{u,d\}$.
Quark field smearing (see below)
at the source and the sink significantly enhances the coupling of the nucleon
creation and destruction operators to the physical ground state.
Still, the time distances between the source and the current insertion $t$
as well as between the current and the sink $t_{\mathrm{f}}-t$
need to be taken sufficiently large to suppress
excited state contributions. 

We find the nucleon smeared-smeared effective masses
to be constant for $t\geq 8a$. It suffices if excited state
effects are much smaller than the statistical errors.
These errors however are expected to be
substantially larger for the disconnected
three-point function than for the nucleon two-point
function. Thus, we
set the time of the current insertion to a smaller
value $t=4a\approx 0.29$~fm.
The method that we apply requires us to
fix $t$, but $t_{\mathrm{f}}$ can be varied.
If the $R^{\mathrm{dis}}$ data were not constant
for $t_{\mathrm{f}}\geq 2t= 8a$ then this would have implied that
our choice of $t$ was too ambitious.
Fortunately, employing sink and source smearing,
we find the asymptotic limit
to be effectively reached for $t_{\mathrm{f}}\geq 5a$
and compute the matrix elements
by fitting the above ratios for
$t_{\mathrm{f}}\geq 6a\approx 0.44$~fm to constants.

As an example, in Fig.~\ref{fig:ratio2} we display
the disconnected ratio for strange valence and current
quark masses as a function of $t_{\mathrm{f}}$ for
smeared-smeared three- over two-point functions
for $40^364$ lattices,
together with this fit result. In addition we show the
corresponding smeared-point ratio that converges towards
the same value, giving us additional confidence that
$t$ was chosen sufficiently large to warrant ground state
dominance within our statistical errors.

Based on Ref.~\cite{Bali:2005fu}
we improve the overlap of our nucleon creation operator
with the ground state by applying
Wuppertal-smearing~\cite{Gusken:1989ad}
\begin{equation}
\label{eq:wuppertal}
\phi^{(n)}_x=\frac{1}{1+6\delta}\left(\phi^{(n-1)}_x+
\delta\sum_{j=\pm 1}^{\pm 3}U_{x,j}\phi^{(n-1)}_{x+a\hat{\boldsymbol{\jmath}}}\right)
\end{equation}
to quark fields $\phi$, where we set $\delta=0.25$ and use
400 iterations.
We replace the spatial links
$U_{x,j}$ above by APE-smeared~\cite{Falcioni:1984ei}
links
\begin{equation}
\label{eq:smear}
U_{x,i}^{(n)}= P_{\mathrm{SU}(3)}\!\left(\!\alpha\,U_{x,i}^{(n-1)}+\sum_{|j|\neq i}
U_{x,j}^{(n-1)}U^{(n-1)}_{x+a\hat{\boldsymbol{\jmath}},i}U^{(n-1)\dagger}_{x+a\hat{\boldsymbol{\imath}},j}\!\right),
\end{equation}
where $i\in\{1,2,3\}, j\in\{\pm 1,\pm 2,\pm 3\}$.
$P_{\mathrm{SU}(3)}$ denotes a projection operator into the $\mathrm{SU}(3)$ group
and the sum is over the four spatial ``staples'', surrounding $U_{x,i}$.
We employ 25 such gauge covariant smearing iterations and use 
the weight factor $\alpha=2.5$. For the projector we
somewhat deviate from
Ref.~\cite{Bali:2005fu} and maximize
$\mathrm{Re}\,\mathrm{Tr}\,[ A^{\dagger} P_{\mathrm{SU}(3)}(A)]$,
iterating over $\mathrm{SU}(2)$ subgroups.
The connected part, for which the statistical accuracy is less of an
issue, is obtained with a less effective smearing
at the larger, fixed value $t_{\mathrm{f}}=15a$, varying $t$.

We stochastically estimate $\mathrm{Tr}_t\,M^{-1}$.
For this purpose we employ
$N$ complex $\mathbb{Z}_2$ noise vectors,
$|\eta^i\rangle_t$, $i=1,\ldots,N$,
whose spacetime $\otimes$ spin $\otimes$ colour
components $e^{i\phi}$ carry uncorrelated
random phases $\phi\in\{\pm\pi/4,\pm 3\pi/4\}$ at the
time $t$ and are set to zero elsewhere, to reduce the noise
(partitioning~\cite{Bernardson:1993yg}).

Solving the linear systems
\begin{equation}
\label{eq:lin}
M|s^i\rangle_t=|\eta^i\rangle_t
\end{equation}
for $|s^i\rangle_t$ we can then
substitute,
\begin{equation}
\label{eq:set}
\mathrm{Tr}_tM_E^{-1}=\frac1N\sum_i^N{}_t\langle\eta^i|s^i\rangle_t
=
\mathrm{Tr}_tM^{-1}+{\mathcal O}\left(\frac{1}{\sqrt{N}}\right)\,.
\end{equation}
The inner product is only taken over three-space,
spin and colour indices.
In the case of the scalar matrix element it is relatively
easy to push the stochastic error below the level
of the inherent error from fluctuations between
gauge configurations\footnote{Note that both error sources
will scale in proportion to $1/\sqrt{n_{\mathrm{conf}}}$.}~\cite{Bali:2009hu}.
Therefore, here we do not need to employ
the Truncated Solver Method (TSM)~\cite{Bali:2009hu} and do not exploit
the hopping parameter
expansion.
Instead, to reduce the dominant gauge error, we compute the
nucleon two-point functions for four equidistant
source times on each gauge configuration. In addition
we exploit backwardly propagating nucleons, replacing
the positive parity
projector $\frac12(\mathds{1}+\gamma_4)$ by
$\frac12(\mathds{1}-\gamma_4)$ within the nucleon
two-point function, $C_{\mathrm{2pt}}$. Consequently, the noise vectors are
seeded on 8 time slices simultaneously, reducing
the degree of time partitioning. We find
this not to have any adverse effect on the stochastic
error. In addition to the 48 
(4 timeslices times 4 spinor components times 3 colours)
point-to-all sources
necessary to compute the
two-point functions we solve for $N=50$ noise vectors
per configuration and current quark mass. 

\section{Renormalization}
\label{sec:renorm}
In the continuum, for light quark flavours $q$,
the $\sigma_q$-terms are invariant under
renormalization
group transformations. However, Wilson fermions explicitly
break chiral symmetry and this enables mixing not
only with gluonic contributions but also with other quark flavours.
Note that due to the use of a quenched strange quark,
the renormalization of the corresponding matrix element
is particularly large and needs to be studied carefully. 
A consistent $\mathcal{O}(a)$ improvement of the
quark scalar matrix elements
requires the inclusion of the gluonic operator $aGG$.
We have not measured this as yet. Therefore
we will neither include any $\mathcal{O}(a)$ improvement
of the renormalization constants nor of the scalar current.
However, we will account for the mixing between quark flavours.

We follow the procedure outlined in Ref.~\cite{Gockeler:2004rp},
see also Sec.~6 of Ref.~\cite{Bhattacharya:1997ht}.
The same result can be obtained by taking derivatives of the
nucleon mass with respect to the sea quark
masses~\cite{Bhattacharya:2005rb} via the Feynman-Hellmann theorem.
This also holds for the case of a partially quenched strange
quark~\cite{Babich:2010at,roger}. In the renormalization the
strangeness matrix element will receive subtractions
not only from light quark disconnected but also from 
the numerically larger connected diagrams. This has
first been pointed out in Ref.~\cite{Michael:2001bv}.

For Wilson actions the vector
Takahashi-Ward identity (VWI) lattice quark mass is given by
\begin{equation}
\label{eq:vwi}
m_q=\frac{1}{2a}\left(\frac{1}{\kappa_q}-\frac{1}{\kappa_{\mathrm{c,sea}}}\right)\,.
\end{equation}
In the partially quenched theory
one can define a
$\kappa_{\mathrm{c,val}}(\beta,\kappa_{\mathrm{sea}})\neq\kappa_{\mathrm{c,sea}}(\beta)$
as the $\kappa$-value at which the valence pseudoscalar mass vanishes. 

We distinguish between singlet and non-singlet quark masses
that will renormalize differently.
In the case of
the theory with $n_{\mathrm{F}}=2+1$ ($m_u=m_d=m_{ud}$) sea quarks,
we have a mass term
\begin{align}
{\mathcal L}_m&=m_{ud}(\bar{u}u+\bar{d}d)+m_s\bar{s}s\nonumber\\&=
\overline{m}\,\bar{\psi}\mathds{1}\psi+m^{ns}\bar{\psi}\lambda_8\psi\,,\\
\overline{m}&=\frac13(2m_{ud}+m_s)\,,
\end{align}
where $\bar{\psi}=(\bar{u},\bar{d},\bar{s})$;
the lattice singlet quark mass $\overline{m}$ is given by the average of the sea quark
masses.

\begin{figure}[t]
\includegraphics[width=.48\textwidth,clip=]{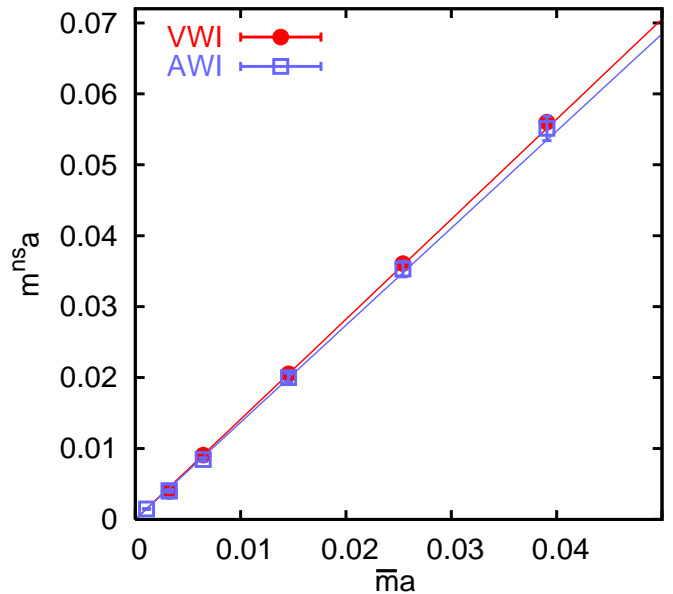}
\caption{Determination of the slope $Z_m^s/Z_m^{ns}=1+\alpha_{\mathrm{Z}}$,
see Eqs.~(\protect\ref{eq:1}) and (\protect\ref{eq:2}).\label{fig:renorm}}
\end{figure}

A renormalized quark mass for the flavour $q$
at a scale $\mu$ is given by
\begin{equation}
\label{eq:ren}
m_q^{\mathrm{ren}}(\mu)=Z_m^s(\mu)\overline{m}+Z_m^{ns}(\mu)(m_q-\overline{m})\,,
\end{equation}
where singlet quark masses renormalize with
a renormalization constant $Z_m^s(\mu)$ and non-singlet combinations
with $Z_m^{ns}(\mu)$.
Notice that the ratio $Z_m^s/Z_m^{ns}$ is independent of the scale
$\mu$ but depends
on the lattice spacing $a$, through the coupling $\alpha_s(a^{-1})$.

In the partially quenched theory with mass degenerate sea quarks
Eq.~(\ref{eq:ren}) results in
\begin{align}
\overline{m}^{\mathrm{ren}}(\mu)&=Z^s_m(\mu) \overline{m}\,,\\
\overline{m}^{\mathrm{ren}}(\mu)-
m^{\mathrm{val,ren}}(\mu)
&=Z^{ns}_m(\mu)(\overline{m}-m^{\mathrm{val}})\,,
\end{align}
where we introduce a VWI valence quark mass through
\begin{equation}
m^{\mathrm{val}}=\frac{1}{2a}
\left(\frac{1}{\kappa_{\mathrm{val}}}-\frac{1}{\kappa_{\mathrm{c,sea}}}\right)\,.
\end{equation}
At $\kappa_{\mathrm{val}}=\kappa_{\mathrm{c,val}}$, $m^{\mathrm{val,ren}}$ vanishes
so that we obtain
\begin{align}
\frac{Z_m^s}{Z_m^{ns}}&=\frac{m^{ns}}{\overline{m}}=\left.\frac{\overline{m}-m^{\mathrm{val}}}{\overline{m}}
\right|_{\kappa_{\mathrm{val}}=\kappa_{\mathrm{c,val}}}\nonumber\\\label{eq:1}
&=
\frac{\kappa_{\mathrm{sea}}^{-1}-\kappa_{\mathrm{c,val}}^{-1}}{
\kappa_{\mathrm{sea}}^{-1}-\kappa_{\mathrm{c,sea}}^{-1}}=:1+\alpha_{\mathrm{Z}}\,.
\end{align}
The non-singlet mass above can also be obtained from the
bare AWI mass,
\begin{equation}
\label{eq:awi}
m^{ns,\mathrm{ren}}=Z_m^{ns}m^{ns}=\frac{Z_A^{ns}}{Z_P^{ns}}m^{\mathrm{AWI}}\,,
\end{equation}
that renormalizes with the ratio of the renormalization constants
$Z_A^{ns}$ over $Z_P^{ns}$
of the non-singlet axial and pseudoscalar currents. This results in the
alternative prescription to Eq.~(\ref{eq:1}),
\begin{equation}
\label{eq:2}
\frac{Z_m^s}{Z_m^{ns}}=\frac{Z_A^{ns}}{Z_m^{ns}Z_P^{ns}}
\frac{2a\,m^{\mathrm{AWI}}}{\kappa_{\mathrm{sea}}^{-1}-\kappa_{\mathrm{c,sea}}^{-1}}\,,
\end{equation}
where $m^{\mathrm{AWI}}$ is calculated at $\kappa_{\mathrm{sea}}$.
The required combination of scalar, pseudoscalar and axial
non-isosinglet
renormalization factors for our simulation with $n_{\mathrm{F}}=2$ sea
quarks at
$\beta=5.29$ reads~\cite{Gockeler:2010yr}
\begin{equation}
Z_A^{ns}/(Z_m^{ns}Z_P^{ns})=0.988\pm 0.031\,.
\end{equation} 
The two methods that differ by terms of $\mathcal{O}(a)$
are illustrated in
Fig.~\ref{fig:renorm} for our configurations with
$n_{\mathrm{F}}=2$ mass degenerate sea quark flavours.

The results read
\begin{equation}
\label{eq:delta}
\alpha_{\mathrm{Z}}=\left\{\begin{array}{ccl}0.411(13)&,&\mathrm{VWI}\\
0.369(22)&,&\mathrm{AWI}\,.\end{array}\right.
\end{equation}
Corrections to these values at a non-zero quark mass $m$
will be of $\mathcal{O}(ma)$;
the differences between the two definitions are indicative of
$\mathcal{O}(a)$ effects.
We remark that $\alpha_{\mathrm{Z}}$ is of leading order $\alpha_s^2$ in the
strong coupling parameter and it will therefore decrease
with the lattice spacing $a$.

For 
$n_{\mathrm{F}}=2+1$ sea quarks
Eq.~(\ref{eq:ren}) amounts to
\begin{align}
\label{eq:ren2}
m_q^{\mathrm{ren}}(\mu)&=Z_m^{ns}(\mu)m_q+(Z^s_m(\mu)-Z_m^{ns}(\mu))
\overline{m}\\\nonumber
&=Z_m^{ns}(\mu)\left(m_q+\alpha_{\mathrm{Z}}\,\overline{m}\right)\,,
\end{align}
where $\alpha_{\mathrm{Z}}$ is defined in Eq.~(\ref{eq:1}).
This can be written as
\begin{widetext}
\begin{equation}
\left(
\begin{array}{c}
m_u(\mu)\\
m_d(\mu)\\
m_s(\mu)\end{array}\right)^{\mathrm{ren}}
=Z^{ns}_m(\mu,a)
\left(
\begin{array}{ccc}
1+\frac{\alpha_{\mathrm{Z}}(a)}{3}&\frac{\alpha_{\mathrm{Z}}(a)}{3}&\frac{\alpha_{\mathrm{Z}}(a)}{3}\\
\frac{\alpha_{\mathrm{Z}}(a)}{3}&1+\frac{\alpha_{\mathrm{Z}}(a)}{3}&\frac{\alpha_{\mathrm{Z}}(a)}{3}\\
\frac{\alpha_{\mathrm{Z}}(a)}{3}&\frac{\alpha_{\mathrm{Z}}(a)}{3}&1+\frac{\alpha_{\mathrm{Z}}(a)}{3}\end{array}\right)
\left(
\begin{array}{c}
m_u(a)\\
m_d(a)\\
m_s(a)\end{array}\right)^{\mathrm{lat}}\,,
\end{equation}
where the lattice quark masses on the right hand side are defined
by the VWI, Eq.~(\ref{eq:vwi}). For clarity we have included
the lattice spacing dependence above, which below we will drop again.
The sum
$\sum_qm_q^{\mathrm{ren}}(\mu)\langle N|\bar{q}q|N\rangle^{\mathrm{ren}}(\mu)$
is invariant under renormalization group transformations\footnote{
Note that these scalar matrix
elements are differences of scalar currents $\bar{q}q$
within the nucleon, relative to their vacuum expectation values. Therefore,
unlike the chiral condensates alone, they do not undergo any
additive renormalization and do not mix with an $a^{-3}\mathds{1}$
term.}.
Therefore the scalar lattice matrix elements will renormalize with the
inverse matrix above:
the different quark contributions will mix in the mass non-degenerate
case.

We now turn to the situation of interest of $n_{\mathrm{F}}=2$ light sea
quarks, with a quenched strange quark. This means that the singlet mass
$\overline{m}^{\mathrm{lat}}=(m_u^{\mathrm{lat}}+m_d^{\mathrm{lat}})/2$ will not depend on the
strange quark mass anymore. (The superscript
``lat'' has been added for clarity.) Eq.~(\ref{eq:ren2}) can now be
written as
\begin{equation}
\label{eq:ren3}
\left(
\begin{array}{c}
m_u(\mu)\\
m_d(\mu)\\
m_s(\mu)\end{array}\right)^{\mathrm{ren}}
=Z^{ns}_m(\mu)
\left(
\begin{array}{ccc}
1+\frac{\alpha_{\mathrm{Z}}}{2}&\frac{\alpha_{\mathrm{Z}}}{2}&0\\
\frac{\alpha_{\mathrm{Z}}}{2}&1+\frac{\alpha_{\mathrm{Z}}}{2}&0\\
\frac{\alpha_{\mathrm{Z}}}{2}&\frac{\alpha_{\mathrm{Z}}}{2}&1\end{array}\right)
\left(
\begin{array}{c}
m_u\\
m_d\\
m_s\end{array}\right)^{\mathrm{lat}}\,,
\end{equation}
where the presence of the strange quark is not felt by the sea quarks.
However, the definition of $m_s$ involves $\kappa_{\mathrm{c,sea}}$.
Inverting the above matrix yields\footnote{Here and occasionally
below we omit specifying the external state (in our case
$|N\rangle$) in cases where identities
between hadronic matrix elements are independent of this state.}
\begin{equation}\label{eq:ren22}
\left(
\begin{array}{c}
\langle\bar{u}u\rangle(\mu)\\
\langle\bar{d}d\rangle(\mu)\\
\langle\bar{s}s\rangle(\mu)\end{array}\right)^{\mathrm{ren}}
=\frac{1}{(1+\alpha_{\mathrm{Z}})Z^{ns}_m(\mu)}
\left(
\begin{array}{ccc}
1+\frac{\alpha_{\mathrm{Z}}}{2}&-\frac{\alpha_{\mathrm{Z}}}{2}&0\\
-\frac{\alpha_{\mathrm{Z}}}{2}&1+\frac{\alpha_{\mathrm{Z}}}{2}&0\\
-\frac{\alpha_{\mathrm{Z}}}{2}&-\frac{\alpha_{\mathrm{Z}}}{2}&1+\alpha_{\mathrm{Z}}\end{array}\right)
\left(
\begin{array}{c}
\langle\bar{u}u\rangle\\
\langle\bar{d}d\rangle\\
\langle\bar{s}s\rangle\end{array}\right)^{\mathrm{lat}}\,.
\end{equation}
For the light quark matrix element (i.e.\ the $\sigma_{\pi\mathrm{N}}$-term)
this means that
\begin{equation}
\frac{m_u^{\mathrm{ren}}(\mu)+m_d^{\mathrm{ren}}(\mu)}{2}
\langle N|\bar{u}u+\bar{d}d|N\rangle^{\mathrm{ren}}(\mu)
=
\frac{m_u^{\mathrm{lat}}+m_d^{\mathrm{lat}}}{2}
\langle N|\bar{u}u+\bar{d}d|N\rangle^{\mathrm{lat}}\,,
\end{equation}
while for the strangeness matrix element we obtain
\begin{equation}
\label{eq:ren4}
\left[m_s\langle N|\bar{s}s|N\rangle\right]^{\mathrm{ren}}
=\left[m_s^{\mathrm{lat}}+\frac{\alpha_{\mathrm{Z}}}{2}\left(m_u^{\mathrm{lat}}
+m_d^{\mathrm{lat}}\right)\right]
\left(
\langle N|\bar{s}s|N\rangle^{\mathrm{lat}}-
\frac{\alpha_{\mathrm{Z}}}{2(1+\alpha_{\mathrm{Z}})}\langle N|\bar{u}u+\bar{d}d|N\rangle^{\mathrm{lat}}
\right)\,.
\end{equation}
\end{widetext}
Again, the lattice strange quark
mass is defined as $m_s^{\mathrm{lat}}=(\kappa_s^{-1}-\kappa_{\mathrm{c,sea}}^{-1})/(2a)$.
The same renormalization pattern can also be derived,
employing the Feynman-Hellmann theorem~\cite{Babich:2010at,roger}.

It is evident from Eq.~(\ref{eq:ren22}) that the so-called $y$-ratio
renormalizes
as follows,
\begin{align}\label{eq:yrat}
y
&:=\frac{2\langle N|\bar{s}s|N\rangle^{\mathrm{ren}}}{\langle N|\bar{u}u+\bar{d}d|N\rangle^{\mathrm{ren}}}\nonumber\\
&=
(1+\alpha_{\mathrm{Z}})\frac{2\langle N|\bar{s}s|N\rangle^{\mathrm{lat}}}{\langle N|\bar{u}u+\bar{d}{d}|N\rangle^{\mathrm{lat}}}-\alpha_{\mathrm{Z}}\,.
\end{align}

\section{Results}
\label{sec:results}
As discussed in Sec.~\ref{sec:method} above we employ three
hopping parameter values,
$\kappa_{ud}=\kappa_{\mathrm{sea}}=0.13632$, $\kappa_m=0.13609$ and
$\kappa_s=0.13550$, that correspond to the pseudoscalar masses
$m_{\mathrm{PS}}\approx 285$~MeV, 450~MeV and 720~MeV, respectively,
see Eqs.~(\ref{eq:mass1}) -- (\ref{eq:mass3}). We use all these
$\kappa$-values for the valence quarks ($\kappa_{\rm val}$)
as well as for the current insertions $\bar{q}q$ ($\kappa_{\mathrm{cur}}$).
This amounts to nine combinations for the disconnected
ratios $R^{\mathrm{dis}}$ while for the connected part
$\kappa_{\mathrm{cur}}=\kappa_{\mathrm{val}}$.

We will explore the dependence of the lattice matrix elements
on the current quark mass and investigate finite size
effects, using a partial summation method. Subsequently,
unrenormalized and renormalized valence and sea quark
contributions are studied. Finally, we compute the
light $\sigma$-term, the mass contributions
$f_{T_s}$ and $f_{T_G}$ and the $y$-ratio.

\subsection{Dependence of the lattice matrix elements on the current
quark mass}
\begin{figure}
\includegraphics[width=.48\textwidth,clip=]{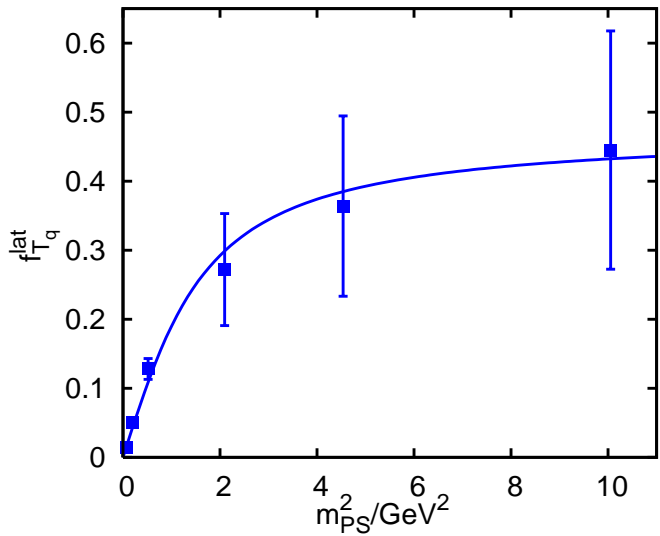}
\caption{The unrenormalized mass fraction $f_{T_q}^{\mathrm{lat}}$,
as a function of the current pseudoscalar mass
on the $L=32a$ lattices, at the smallest valence mass
$m_{\mathrm{PS}}\approx 285$~MeV.\label{fig:limiting}}
\end{figure}

From the heavy quark expansion it is evident that
$f_{T_q}\propto\langle N|GG|N\rangle/m_{\mathrm{N}}\propto f_{T_G}$ for
$m_q\rightarrow \infty$. This can also be seen on the lattice
where the leading non-vanishing
contribution in the hopping parameter expansion is proportional
to the plaquette. To confirm this expected saturation
we compute the scalar matrix element for three additional
current quark masses up to and above the charm quark mass $m_c$
on a subset of 576 $32^364$ lattices, using
our smallest valence quark mass. We remark that
for an $\mathcal{O}(a)$ improvement of the operator $\bar{q}q$,
mixing with $aGG$ needs to be considered. These improvement terms
will become large if $m_qa$ is big. In fact the
renormalized $f_{T_q}$ becomes
negative at about twice the strange quark mass if we neglect
such effects.
We have not investigated gluonic contributions as yet
and therefore the behaviour at large masses is beyond
the scope of the present study.
For very small quark masses we would expect
$f_{T_q}\propto m_q$. Indeed, these expectations are
confirmed for the unrenormalized $f_{T_q}^{\mathrm{lat}}$,
as can be seen in Fig.~\ref{fig:limiting}. Note that the third
data point from the left corresponds to the strange quark mass while
the charm quark can be found at the value
$m_{\mathrm{PS}}^2\approx 8.9\,\mathrm{GeV}^2$.
The arctan-curve is drawn to guide the eye.

\subsection{Finite size effects: partial sums}
\begin{figure}
\includegraphics[height=.48\textwidth,angle=270,clip=]{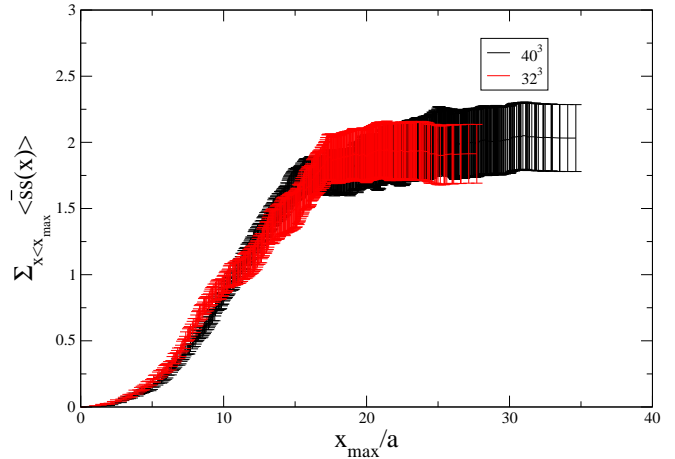}
\caption{Eq.~(\ref{eq:rati}), partially summed up to
a maximum spatial distance from the source $x_{\max}$, see
Eq.~(\protect\ref{eq:partial}).
\label{fig:partial}}
\end{figure}

In order to investigate finite size effects
we find it worthwhile to replace
the numerator of Eq.~(\ref{eq:rati}) by a partial sum:
\begin{equation}
\label{eq:partial}
-\sum_{|{\mathbf x}|\leq x_{\rm max}}\frac{
\left\langle
C_{\rm 2pt}(t_{\mathrm{f}})
\,
\mathrm{Tr}\left[M^{-1}({\mathbf x},t;{\mathbf x},t)\right]
\right\rangle_{\!c}}{\langle C_{\mathrm{2pt}}(t_{\mathrm{f}})\rangle}\,.
\end{equation}
The subscript $c$ (connected part) indicates
that we subtract the
product of the two individual vacuum expectation values from the
numerator.
Since no zero momentum projection is performed at the source
of the two-point function,
that resides at the spatial position $\mathbf{x}=\mathbf{0}$,
the result will depend on the cut-off 
$x_{\max}$.
We expect the summand
at large $|\mathbf{x}|$ to fall off exponentially
in $|\mathbf{x}|$ so that these values 
will eventually not contribute to the signal
anymore but just increase the statistical noise.
At very large spatial volumes one
may therefore consider to perform such a partial sum only,
thereby reducing the statistical error,
and to estimate the induced bias by parameterizing the asymptotic
fall-off.

\begin{table}
\caption{The disconnected contribution to the
scalar lattice matrix elements for different
$\kappa$-combinations.\label{tab:scalar}}
\begin{center}
\begin{ruledtabular}
\begin{tabular}{cccc}
$\kappa_{\mathrm{val}}$&$\kappa_{\mathrm{cur}}$&$V$&$\langle N|\bar{q}q|N\rangle^{\mathrm{lat}}_{\mathrm{dis}}$\\\hline
       &0.13550&$32^364$&2.01(21)\\
       &       &$40^364$&2.17(25)\\
0.13550&0.13609&$32^364$&2.27(22)\\
       &       &$40^364$&2.43(27)\\
       &0.13632&$32^364$&2.38(23)\\
       &       &$40^364$&2.55(29)\\\hline
       &0.13550&$32^364$&1.97(20)\\
       &       &$40^364$&2.06(25)\\
0.13609&0.13609&$32^364$&2.19(22)\\
       &       &$40^364$&2.28(26)\\
       &0.13632&$32^364$&2.21(23)\\
       &       &$40^364$&2.36(28)\\\hline
       &0.13550&$32^364$&1.96(23)\\
       &       &$40^364$&1.93(27)\\
0.13632&0.13609&$32^364$&2.06(24)\\
       &       &$40^364$&2.05(29)\\
       &0.13632&$32^364$&1.67(26)\\
       &       &$40^364$&1.86(31)
\end{tabular}
\end{ruledtabular}
\end{center}
\end{table}

We display the partial sums for the $L=32a$ and $L=40a$
lattices for the strangeness current at $\kappa_{\mathrm{val}}=
\kappa_{\mathrm{sea}}$
in Fig.~\ref{fig:partial}. 
We do not detect any statistically significant dependence
of the curves on the value
of $t_{\mathrm{f}}$ and show the results obtained
at $t_{\mathrm{f}}=6a$.
At small $x_{\max}$ we see
the naively expected $x_{\max}^3$ volume scaling.
This becomes flatter
around $x_{\max}\approx 8a$ but only saturates to a constant
once the boundary of the $L=32a$ box is hit at
$x_{\max}=16a$. Increasing $x_{\max}$ beyond this
value means that in the case of this smaller lattice only the
lattice ``corners'' are summed up.
However, the $L=40a$ data saturate at the same distance,
rather than at $20a$, indicating that indeed the nucleon
is well accommodated within this box size and that finite size
effects are small.

\subsection{Sea and valence quark contributions}
\begin{figure}
\includegraphics[width=.48\textwidth,clip=]{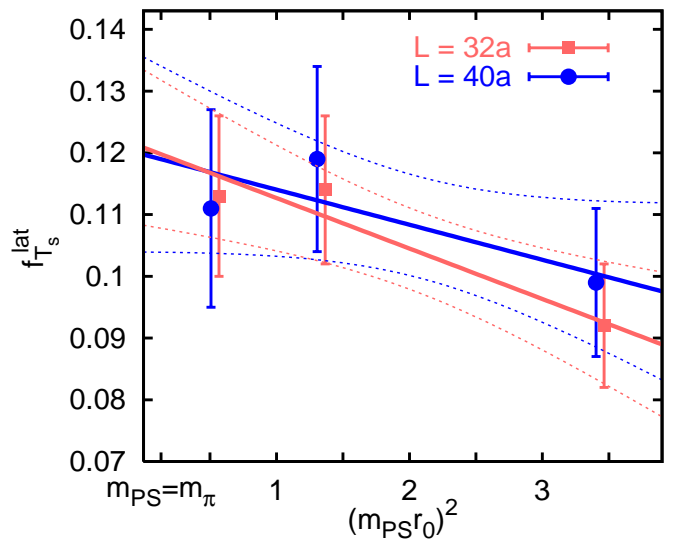}
\caption{The unrenormalized strange quark mass fraction $f_{T_s}^{\mathrm{lat}}$,
as a function of the valence pseudoscalar mass, for the two volumes.\label{fig:ftschiral}}
\end{figure}

The results of the bare disconnected scalar matrix elements
for the two volumes (and three
quark masses $m_q\leq m_s$) are displayed in Table~\ref{tab:scalar}.
For disconnected terms
$\kappa_{\mathrm{cur}}$
can differ not only from $\kappa_{\mathrm{sea}}$ but also
from the $\kappa_{\mathrm{val}}$ of the nucleon's
valence quarks. In Fig.~\ref{fig:ftschiral} we display the dependence
of the unrenormalized $f_{T_s}^{\mathrm{lat}}$-values on the valence quark
mass of the proton for both volumes, together with linear
chiral extrapolations. The right-most data points
correspond to the strange quark mass and the
left-most points to the present sea quark mass.
The volume dependence is not significant
and neither are the differences between the values obtained at
the smallest mass point and the chirally extrapolated numbers.
The results need
to be renormalized and this is possible at
$\kappa_{\mathrm{val}}=\kappa_{\mathrm{sea}}=0.13632$.

To enable the calculation of the light $\sigma$-term and
the renormalization of the strangeness matrix element, we
also compute the connected contribution for
$\kappa_{\mathrm{cur}}=\kappa_{\mathrm{val}}=\kappa_{\mathrm{sea}}=0.13632$,
using the traditional sequential propagator method. 
We obtain
$\langle N|\bar{u}u+\bar{d}d|N\rangle^{\mathrm{lat}}_{\mathrm{con}}
=8.43(73)$ and 8.35(43)
for the $L=32a$ and $L=40a$ lattices, respectively.
This means that at the pseudoscalar mass $m_{\mathrm{PS}}\approx 285$~MeV
the relative contribution of the disconnected matrix 
element reads
\begin{equation}\label{eq:rr}
r^{\mathrm{lat}}
=\frac{\langle N|\bar{u}u+\bar{d}d|N\rangle_{\mathrm{dis}}^{\mathrm{lat}}}
{\langle N|\bar{u}u+\bar{d}d|N\rangle^{\mathrm{lat}}}=
\left\{\begin{array}{l}0.284(36)\,\,,\,L=32a\\
0.308(37)\,\,,\,L=40a\,.\end{array}\right.
\end{equation}
The unrenormalized values of Table~\ref{tab:scalar}
seem to be fairly independent of the current quark mass.
The ratio of the strangeness matrix element over a
light sea quark contribution undergoes the renormalization
\begin{equation}
\frac{2\langle \bar{s}s\rangle^{\mathrm{ren}}}
{\langle \bar{u}u+\bar{d}d\rangle_{\mathrm{dis}}^{\mathrm{ren}}}
=\frac{2\langle \bar{s}s\rangle^{\mathrm{lat}}-
\frac{\alpha_{\mathrm{Z}}}{1+\alpha_{\mathrm{Z}}}\langle \bar{u}u+\bar{d}d\rangle^{\mathrm{lat}}}
{\langle \bar{u}u+\bar{d}d\rangle_{\mathrm{dis}}^{\mathrm{lat}}-
\frac{\alpha_{\mathrm{Z}}}{1+\alpha_{\mathrm{Z}}}
\langle \bar{u}u+\bar{d}d\rangle^{\mathrm{lat}}}\,.
\end{equation}
The renormalization of the numerator is obtained from Eq.~(\ref{eq:ren22})
while the denominator can be split into
two parts that renormalize with $Z_m^s$ and with $Z_m^{ns}$,
respectively:
$\langle \bar{u}u+\bar{d}d\rangle
-\langle \bar{u}u+\bar{d}d\rangle_{\mathrm{con}}$.
The $\mathrm{SU(3)_F}$ flavour symmetry of the
unrenormalized sea obviously cannot disappear when subtracting
the same (large) terms from
the numerator and the denominator. However, this subtraction
results in large statistical uncertainties. For instance on the large
volume the above ratio reads $1.7\pm 5.5$ with an additional systematic
uncertainty of $0.5$ from the value of the renormalization parameter
$\alpha_{\mathrm{Z}}$, Eq.~(\ref{eq:delta}). 
We conclude that the renormalized
sea is $\mathrm{SU(3)_F}$ symmetric within a factor of about
five.

The disconnected fraction $r$ of Eq.~(\ref{eq:rr}) will
undergo the renormalization
\begin{equation}
\label{eq:rren}
r^{\mathrm{ren}}=(1+\alpha_{\mathrm{Z}})r^{\mathrm{lat}}-\alpha_{\mathrm{Z}}
=0.024(5)^{+29}_{-9}\,.
\end{equation}
The value quoted is obtained on the $L=40a$ volume using the 
VWI prescription, with a systematic
error that incorporates the difference between the two determinations
of the renormalization constant ratios
Eq.~(\ref{eq:delta})
and their respective uncertainties.
For the renormalized $y$-parameter that is defined in
Eq.~(\ref{eq:yrat}) this
implies that
\begin{equation}
y=r^{\mathrm{ren}}\frac{2\langle \bar{s}s\rangle^{\mathrm{ren}}}
{\langle \bar{u}u+\bar{d}d\rangle_{\mathrm{dis}}^{\mathrm{ren}}}
\approx r^{\mathrm{ren}}\,,
\end{equation}
where the approximation holds within a factor of five, see also
Eq.~(\ref{eq:yresult}) below.

\subsection{The light and strange $\boldsymbol{\sigma}$-terms}
Combining all information results in the renormalized values
for the pion-nucleon $\sigma$-term at the simulated sea quark mass
\begin{align}\label{eq:sigres}
\sigma_{\mathrm{PS\,N}}
&=\left\{\begin{array}{c}0.0378(39) a^{-1}=0.264(26)(2) r_0^{-1}\,,\,L=32a\\
0.0389(36)a^{-1}=0.272(25)(2)r_0^{-1}\,,\,L=40a\end{array}\right.
\end{align}
at the two different volumes, where the second error
is due to the uncertainty of the chirally extrapolated $r_0$-value.
The above $\sigma$-term can also be obtained from the derivative
of the nucleon mass with respect to the logarithm of the light quark
mass. Using the fact that at small $m_{\mathrm{PS}}$, $m_u+m_d\propto
m_{\mathrm{PS}}^2$, we can write
\begin{equation} 
\label{eq:siglo}
\sigma_{\mathrm{PS\,N}}=
m_u\frac{\partial m_{\mathrm{N}}}{\partial m_u}+
m_d\frac{\partial m_{\mathrm{N}}}{\partial m_d}
\approx
m_{\mathrm{PS}}^2\frac{d m_{\mathrm{N}}}{d m_{\mathrm{PS}}^2}\,.
\end{equation}

\begin{figure}
\includegraphics[width=.48\textwidth,clip=]{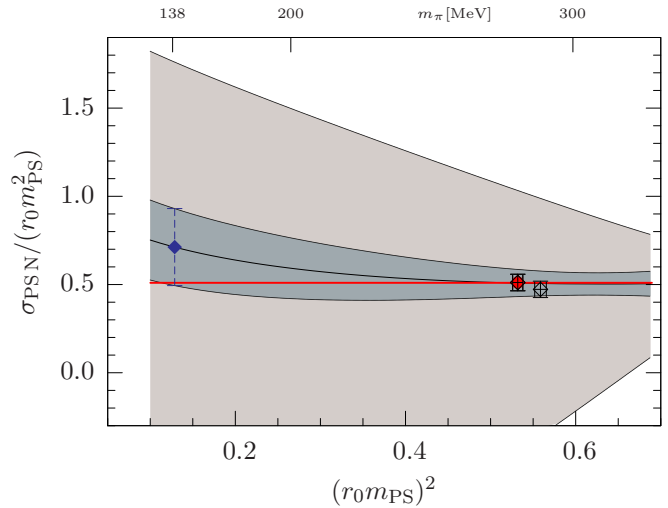}
\caption{Extrapolation of $\sigma_{\mathrm{PS\,N}}/m_{\mathrm{PS}}^2$
to the physical point~\protect\cite{sternbeck} using covariant B$\chi$PT.
The open symbols correspond to the values that we directly
obtain at $m_{\mathrm{PS}}r_0\approx 0.73$ on the
$L=40a$ volume (left) and for $L=32a$ (right).
The broad error band is obtained
when ignoring this constraint. The horizontal line
denotes the (constant) leading order expectation.\label{fig:sigma}}
\end{figure}

To leading order in chiral perturbation theory
$dm_{\mathrm{N}}/d{m_{\mathrm{PS}}^2}=\mathrm{const}$.
This linear assumption suggests to multiply
the result Eq.~(\ref{eq:sigres}) 
with the ratio $m_{\pi,\mathrm{phys}}^2/m_{\mathrm{PS}}^2$ to
obtain the physical $\sigma$-term
$\sigma^{{\mathrm{phys,}0}}_{\pi\mathrm{N}}=0.064(6)r_0^{-1}=25(3)(1)$~MeV.
In general, however, higher order corrections will lead to some
curvature.

\begin{table}
\caption{QCDSF pseudoscalar and nucleon masses~\protect\cite{sternbeck}
at $\beta=5.29$ ($a^{-1}\approx 2.7\,\,\mathrm{GeV}$) and
$\beta=5.40$ ($a^{-1}\approx 3.2\,\,\mathrm{GeV}$).\label{tab:nucleon}}
\begin{center}
\begin{ruledtabular}
\begin{tabular}{cccccccc}
$\beta$ & $\kappa$ & $V$ &   $am_{\mathrm{PS}}$ &$am_{\mathrm{N}}$ &$r_0m_{\mathrm{PS}}$  &   $r_0M_{\mathrm{N}}$\\\hline
5.29 & 0.13620 & $24^348$ &   0.1552(6) &   0.467(5) &  1.084(9) &   3.26(4)\\
5.29 & 0.13632 & $24^348$ &   0.1112(9) &   0.425(6) &  0.776(8) &   2.97(5)\\
5.29 & 0.13632 & $32^364$ &   0.1070(4) &   0.390(5) &  0.747(6) &   2.72(4)\\
5.29 & 0.13632 & $40^364$ &   0.1050(3) &   0.381(3) &  0.733(6) &   2.66(3)\\
5.40 & 0.13660 & $32^364$ &   0.0845(6) &   0.353(7) &  0.700(8) &   2.92(7)\\
5.40 & 0.13660 & $48^364$ &   0.0797(3) &   0.314(5) &  0.660(7) &   2.60(5)
\end{tabular}
\end{ruledtabular}
\end{center}
\end{table}

Fortunately, we do not only know the $\sigma$-term at
$\kappa=0.13632$ but also the nucleon mass~\cite{sternbeck} at
other values of $\kappa_{\mathrm{val}}=\kappa_{\mathrm{sea}}$, at
$\beta=5.29$ ($a^{-1}\approx 2.71\,\,\mathrm{GeV}$)
and at $\beta=5.4$
($a^{-1}\approx 3.22\,\,\mathrm{GeV}$), see
Table~\ref{tab:nucleon}.
A combined $\mathcal{O}(p^4)$ covariant B$\chi$PT~\cite{Becher:2001hv}
fit to these data within the window
$250\,\mathrm{MeV}< m_{\mathrm{PS}}<
430\,\mathrm{MeV}$, imposing
our directly obtained value
of $\sigma_{\mathrm{PS\,N}}$ as an additional
constraint, results in the preliminary number~\cite{sternbeck}
\begin{equation}
\sigma^{\mathrm{phys}}_{\pi\mathrm{N}}=(38\pm 12)\,\mathrm{MeV}
\end{equation}
at the physical point. The error includes both
the statistical uncertainty of the fit
and the systematics from varying the low energy
parameters $c_2$, $c_3$ and
$l_3$ within their phenomenologically
allowed ranges~\cite{Becher:2001hv,Frink:2004ic,flag}.
A detailed analysis will be presented in
Ref.~\cite{sternbeck}.
We display the result of this extrapolation in
Fig.~\ref{fig:sigma} for the ratio
$\sigma_{\mathrm{PS\,N}}/m^2_{\mathrm{PS}}$ in units of $r_0$,
together with our direct determinations. The broad error band indicates
the result of the same fit, without using our constraint
at $m_{\mathrm{PS}}\approx 285$~MeV.

We now use Eq.~(\ref{eq:ren4}) with $\alpha_{\mathrm{Z}}$ given in
Eq.~(\ref{eq:delta}) to obtain the renormalized strangeness matrix
element from the values given above.
This amounts to subtracting numbers of similar sizes
from each other. There is no noticeable finite size
effect between the $32^3$ and $40^3$ volumes.
For our simulation
point at a low pion mass $m_{\mathrm{PS}}\approx 285$~MeV we obtain
the values,
$a[m_s\langle N|\bar{s}s|N\rangle]^{\mathrm{ren}}=
0.005(6)$ and 0.008(6) for the two determinations
of the renormalization
parameter $\alpha_{\mathrm{Z}}$ from the VWI and AWI, respectively.
 
Of particular phenomenological interest is the dimensionless
strange quark contribution to the nucleon mass
\begin{equation}
f_{T_s}=\frac{[m_s\langle N|\bar{s}s|N\rangle]^{\mathrm{ren}}}{m_{\mathrm{N}}}
=0.012(14)^{+10}_{-3}\,.
\end{equation}
Again, we quote the value obtained from the VWI prescription, with a systematic
error that incorporates the difference between the two determinations
of the renormalization constant ratios and their respective uncertainties.
This may be indicative
of $\mathcal{O}(a)$ effects.
The problem of large cancellations cannot be overcome
easily.
One needs to get closer to the continuum limit so that
$\alpha_{\mathrm{Z}}$ approaches zero.
For instance, at $\beta=5.40$, $\alpha_{\mathrm{Z}}\approx 0.2$~\cite{Gockeler:2004rp},
significantly reducing the subtraction of the connected
diagram (and probably the value of $\langle N|\bar{s}s|N\rangle^{\mathrm{lat}}$ that
will contain a smaller light quark contribution).

The result obtained is interesting insofar as it
suggests a scalar strangeness of less than $4\,\%$ of the nucleon mass,
$\sigma_s=12^{+23}_{-16}$~MeV. In spite of the relative enhancement
by the ratio $m_s/m_{ud}> 25$ this is not bigger
than the pion-nucleon $\sigma$-term above.
This is quite consistent with the finding
of Eq.~(\ref{eq:rren}) of
a tiny renormalized light sea quark
participation in $\sigma_{\mathrm{PS\,N}}$.
We remark that
taking the combination $m_s^{\mathrm{lat}}\langle N| \bar{s}s|N\rangle^{\mathrm{lat}}$
without the proper subtraction would have resulted in
$f_{T_s}\approx 0.12$, even bigger than the
light quark mass contribution of about 0.09, at our
light quark mass value that exceeds the physical one by a factor of
about four. Neglecting the mixing with light quarks in the
renormalization is probably the main reason why this contribution
was overestimated in the pioneering lattice studies, see e.g.\
Ref.~\cite{Gusken:1999te}
and references therein. Early results are also summarized in
Ref.~\cite{Young:2009ps}

We can constrain the scale-independent
$y$-ratio of Eq.~(\ref{eq:yrat}),
\begin{align}
y&=
\left\{\begin{array}{c}
(1+\alpha_{\mathrm{Z}})\,0.333(36)-\alpha_{\mathrm{Z}}\,,\,L=32a\\
(1+\alpha_{\mathrm{Z}})\,0.320(33)-\alpha_{\mathrm{Z}}\,,\,L=40a\end{array}\right.\\
&=\left\{\begin{array}{l}
0.059(37)(28)\,,\,L=32a\\
0.041(37)(29)\,,\,L=40a\,,\end{array}\right.\label{eq:yresult}
\end{align}
where the errors are statistical and the difference
between the two determinations of $\alpha_{\mathrm{Z}}$, respectively.
Again, as the central value, we have taken the result from the VWI
renormalization factor.
From our determination of the pion-nucleon $\sigma$-term we know that the
denominator of Eq.~(\ref{eq:yrat}) will increase by a factor
1.4--1.5 when extrapolated
to the chiral limit.
Based on the weak observed dependence
of $\langle N|\bar{s}s|N\rangle^{\mathrm{lat}}$
on the valence quark mass, see Fig.~\ref{fig:ftschiral},
we would expect the numerator to exhibit
a less pronounced quark mass dependence. Thus a $95~\%$
confidence-level upper limit on
the $y$-parameter $y<0.14$
should also apply at physically light sea quark masses.

Finally, we also predict the gluonic and heavy sea quark
contribution $f_{T_G}$ of Eq.~(\ref{eq:ftg}),
\begin{equation}
f_{T_G}=1-\frac{\sigma_{\pi\mathrm{N}}+\sigma_s}{m_{\mathrm{N}}}
=0.951^{+20}_{-27}\,.
\end{equation}
This means that the light and strange quark flavours
contribute a fraction between 3~\% and 8~\% to the
nucleon mass.

\section{Summary}
\label{sec:outlook}
We directly calculate the light quark and strangeness
$\sigma$-terms on lattices with spatial extents
up to $Lm_{\mathrm{PS}}\approx 4.2$, a lattice spacing
$a^{-1}\approx 2.71$~GeV and a
pseudoscalar mass $m_{\mathrm{PS}}\approx 285$~MeV.
At this mass point and lattice spacing the quark line
disconnected contribution amounts to a fraction
of $r^{lat}\approx 0.3$
of the full result. After renormalization however
we find this number to drop below the 5~\% level, see
Eq.~(\ref{eq:rren}).

At our fixed mass point we obtain the renormalized values,
$\sigma_s=12^{+23}_{-16}$~MeV
and $\sigma_{\mathrm{PS\,N}}=106(11)(3)$~MeV, for the
strangeness and light quark $\sigma$-terms.
Assuming the latter value to depend linearly on $m_{\mathrm{PS}}^2$,
as predicted by leading order chiral perturbation theory,
this corresponds to $25(3)\,\mathrm{MeV}$ at the physical point.
However, from nucleon mass data obtained at
pseudoscalar masses, $250\,\mathrm{MeV}< m_{\mathrm{PS}}<
430\,\mathrm{MeV}$, it is clear that there exists a non-vanishing
curvature. Our direct determination can be used
to strongly constrain an $\mathcal{O}(p^4)$ covariant
baryon chiral perturbation
theory extrapolation of the nucleon mass~\cite{sternbeck}.
The combined fit yields the preliminary value
$\sigma^{\mathrm{phys}}_{\pi\mathrm{N}}=\left(38\pm 12\right)\,\mathrm{MeV}$
at the physical
point. Without the direct information on the slope
at one point the statistical and systematic uncertainties
would have been much larger, see Fig.~\ref{fig:sigma}.
It would be difficult to significantly
reduce this large error, without nucleon mass data at physical and,
possibly, smaller than physical quark masses.

We are also able to exclude values $y>0.14$ of the
$y$-parameter, with a confidence level of 95~\%.
This means that the strangeness contribution to the scalar
coupling of the nucleon is much smaller than that
due to the light quark
$\sigma$-term. To determine the strangeness
$\sigma$-term using the indirect method, i.e.\ the
Feynman-Hellmann theorem, requires $n_{\mathrm{F}}=2+1$
sea quark flavours. Even then
the dependence of
the nucleon mass on the strange quark mass will be very weak and the
tiny slope (and its error) will be amplified by the
mass ratio $m_s/m_{ud}>25$. Therefore, for an accurate
prediction of $f_{T_s}$, and in particular for a non-vanishing lower bound
on its value,
an additional direct determination at one or a few mass points
will be crucial.

We remark that the results presented here have not been
extrapolated to the continuum limit. Neither has the effect of
quenching the strange quark been addressed, except within the
renormalization of the strangeness $\sigma$-term.
\begin{acknowledgments}
We thank Peter Bruns and Ludwig Greil for discussion.
This work was supported by the European Union
under Grant Agreement number 238353 (ITN STRONGnet)
and by the Deutsche
Forschungsgemeinschaft SFB/Transregio 55. Sara Collins
acknowledges support from the Claussen-Simon-Foundation (Stifterverband
f\"ur die Deutsche Wissenschaft).
Andr\'e Sternbeck was supported by the EU
International Reintegration Grant (IRG) 256594.
James Zanotti was supported by the Australian Research Council
under grant FT100100005 and previously
by the Science \& Technology
Facilities Council under grant ST/F009658/1.
Computations were performed on the
SFB/TR55 QPACE supercomputers,
the BlueGene/P (JuGene) and the Nehalem Cluster (JuRoPA) of the
J\"ulich Supercomputer Center, the
IBM BlueGene/L at the EPCC (Edinburgh),
the SGI Altix ICE machines at HLRN (Berlin/Hannover)
and Regensburg's Athene HPC cluster. We thank the
support staffs of these institutions.
The Chroma software suite~\cite{Edwards:2004sx} was used extensively
in this work and gauge configurations
were generated using the BQCD code~\cite{Nakamura:2010qh}
on QPACE and BlueGenes.
\end{acknowledgments}

\end{document}